# LIINUS/SERPIL: a design study for interferometric imaging spectroscopy at the LBT

C. Gál*[a], F. Müller-Sánchez**[b], A. Krabbe[a], F. Eisenhauer[b], C. Iserlohe[a], M. Haug[b], T. M. Herbst[c]
[a]University of Cologne, I. Physics Institute, Zülpicher Str. 77, D-50937 Köln
[b]Max-Plank Institute for extraterrestrial Physics, Giessenbachstr., P.O.Box 1312, D-85741 Garching,
[c]Max-Planck Institute for Astronomy, Königstuhl 17, D-69117 Heidelberg

**ABSTRACT**

LIINUS/SERPIL is a design study to augment LBTs interferometric beam combiner camera LINC-NIRVANA with imaging spectroscopy. The FWHM of the interferometric main beam at 1.5 micron will be about 10 mas, offering unique imaging and spectroscopic capabilities well beyond the angular resolution of current 8-10m telescopes. At 10 mas angular scale, e.g., one resolution element at the distance of the Galactic Center corresponds to the average diameter of the Pluto orbit (79 AU), hence the size of the solar system. Taking advantage of the LBT interferometric beam with an equivalent maximum diameter of 23 m, LIINUS/SERPIL is an ideal precursor instrument for (imaging) spectrographs at extremely large full aperture telescopes. LIINUS/SERPIL will be built upon the LINC-NIRVANA hardware and LIINUS/SERPIL could potentially be developed on a rather short timescale. The study investigates several concepts for the optical as well as for the mechanical design. We present the scientific promises of such an instrument together with the current status of the design study.

## 1  INTRODUCTION

The LBT (**L**arge **B**inocular **T**elescope), located on Mt. Graham near Tucson/Arizona at an altitude of about 3200m, is an innovative project being undertaken by institutions from Europe and USA. The structure of the telescope incorporates two 8.4-meter mirrors on a 14.4 meter center-to-center common mount. This configuration provides the equivalent collecting area of a 12 meter telescope, and when combined coherently, the two optical paths offer diverse possibilities for interferometry. Currently two different beam combiners are being developed for the LBT: the LBTI (**L**arge **B**inocular **T**elescope **I**nterferometer [1]) performing nulling interferometry, and LINC-NIRVANA (the **L**BT **IN**terferometric **C**amera, **N**ear-**IR** / **V**isible **A**daptive i**N**terferometer for **A**stronomy [2]), a Fizeau interferometer incorporating multi-conjugate adaptive optics.

LIINUS (**L**INC-NIRVANA **I**nterferometric **I**maging **N**ear-infrared **U**pgrade **S**pectrograph) is a design study for a near-infrared integral field spectrograph for interferometric observations at the LBT initiated by the I. Physics Institute of the University of Cologne. A similar project SERPIL (**S**pectrograph for **E**nhanced **R**esolution **P**erforming **I**maging interferometry on the **L**BT) was independently launched at the Max-Planck-Institute for Extraterrestrial Physics (MPE) in Garching. The joint cooperation between both groups is reflected in the acronym LIINUS/SERPIL. The instrument under study offers the exceptional opportunity of diffraction limited imaging of an equivalent 23 m telescope, when the light of the two primary mirrors of the LBT is combined coherently in the Fizeau interferometric mode. LIINUS/SERPIL will expand the capabilities of the LINC/NIRVANA camera by adding an integral field spectrographic mode and by providing high resolution infrared spectra of more than 1000 image pixels in a two dimensional field. The design study of LIINUS/SERPIL explores potential technical solutions for diffraction limited interferometric imaging spectroscopy at the LBT.

*gal@ph1.uni-koeln.de; phone +49 221 470-7791; fax +49 221 470-6727; http://www.ph1.uni-koeln.de
** frankmueller@mpe.mpg.de; phone +49 089 30000-3587; fax +49 089 30000-3569; http://www.mpe.mpg.de

## 2 SCIENCE MOTIVATION

The pioneering capabilities of LIINUS/SERPIL will facilitate the investigations of physical processes in many interesting classes of objects. Imaging spectroscopy at 10 to 20 mas angular resolution from J to K-band will provide opportunities for targets in the Solar System, within our own Galaxy, as well as in nearby and distant galaxies. In the following subsections, we summarize the most important science drivers for this instrument.

### 2.1 Nearby Active Galactic Nuclei

The coexistence of an active galactic nucleus (AGN), presumably a black hole with an accretion disk, and a starburst region on scales of a few parsecs around the nucleus, is one of the key issues to investigate in the context of AGN. Increasing evidence that starbursts do occur in the vicinity of AGN [3] has revived the importance of disentangling how the gas is driven to the inner regions of AGN leading to star formation and how AGN and star formation activity impact on each other. LIINUS/SERPIL will be ideally suited for studying the nuclear environments of active galaxies at near-infrared wavelengths (where extinction is a factor of 10 smaller than at optical wavelengths) and with adaptive optics allowing for diffraction-limited angular resolution. At an angular resolution of 20 mas, LIINUS/SERPIL will be able to resolve the central stellar cluster and the molecular torus on scales of one parsec at an object's distance of 10 Mpc. Spectral synthesis models can then be used to infer the age and stellar mass function in individual star-forming regions. In addition, observations at this resolution will allow us to directly test the unified model by separating spatially and spectrally the narrow line clouds from the inner broad line clouds.

Observing in K-band provides us with diagnostics e.g. for stellar content and dynamics (CO bandheads along 2.3 microns), warm molecular gas (rotational-vibrational transitions of molecular hydrogen), star formation rate (e.g. Brγ emission at 2.17 micron) and high excitation processes (coronal lines of [SiVI] at 1.96 micron or [CaVIII] at 2.32 micron). Diffraction limited integral field spectroscopy has already been used to characterize the stellar and gaseous components in the nucleus of several nearby AGNs, including the Circinus galaxy (Fig. 1, [4]). We find vigorous star formation and resolve the stellar component on scales of a few parsecs and observe the presence of young stellar clusters of an average mass of 10000 solar masses. Since the scales on which the gas and stars exist are similar, they suggest that the torus is forming stars and most likely clumpy. Finally, an extensive analysis of the coronal lines showed that they are composite: one narrow component associated to the AGN and a broad component, which is flowing out from the nucleus and which is most probably excited by shocks.

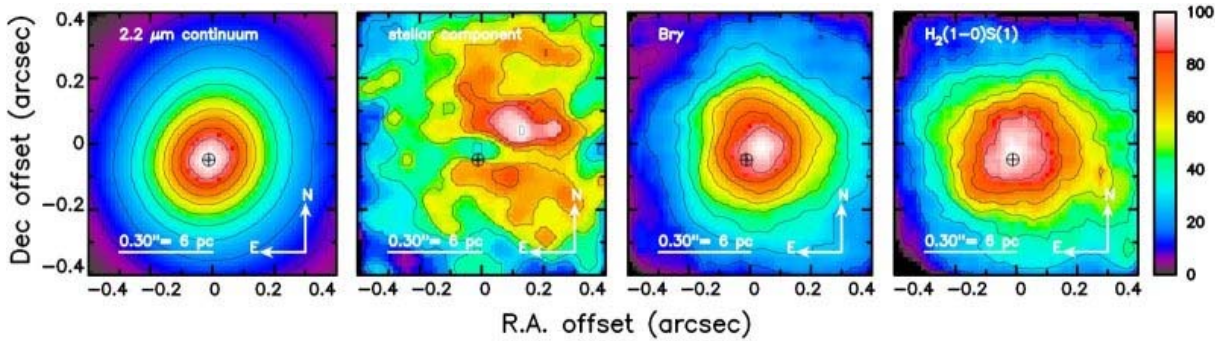

Fig. 1. Intensity Images of the Circinus Galaxy extracted from the SINFONI data cube. Left to right: 2.2 micron continuum, Stellar continuum (flux contained in the stellar absorption bandhead $^{12}CO(2-0)$), Brγ, $H_2$ 1-0S(1).

Integral field spectroscopy is extremely well suited for studying the inner dynamics in nearby AGN's and therefore the determination of the nuclear mass, since it helps to avoid ambiguous interpretations from e.g. slit spectroscopy. The determination of black hole masses in AGN based on stellar dynamics is crucial not only for the verification of the M-sigma relation. It is also important for the assessment of the M-sigma scatter of these galaxies, and to provide a comparison to reverberation masses which might then allow to constrain the size and geometry of the broad line region. The high spatial resolution and integral field capability of LIINUS/SERPIL will provide an ideal combination to do this. Recently, by means of SINFONI at the VLT, [3] have successfully derived the black hole mass in NGC3227 from stellar kinematics for the first time for a Seyfert 1 galaxy.

## 2.2 Stellar content and supermassive black holes in nearby galaxies

Using the diffraction limited angular resolution of LIINUS/SERPIL, the spatially resolved near-IR properties of individual stars in the bulges of nearby galaxies can be investigated. Abundance studies can be carried out by analyzing stellar absorption lines blueward of the 2.3 microns CO bandheads, such as Mg, Si and Ca species allowing to constrain the formation history of galaxy bulges such as M31 or M32.

There is now strong evidence for the presence of massive black holes in the nuclei of a number of nearby galaxies. In these "normal" galaxies, the black hole is starving from the depletion of gas inflow. It seems now established that the Milky Way, NGC4258, and likely M31 each host a supermassive black hole [5]. The mass of the black hole correlates with the bulge velocity dispersion, the so-called M-$\sigma$ relation [6]. LIINUS/SERPIL will help to test the M-$\sigma$ relation in a variety of nearby galaxies. This requires sensitive spectroscopy of the stellar and gas kinematics at high spatial resolution to probe within the radius of influence of the central black hole (see Fig. 2 as an example of NGC1275 observed by [7]) Radii of influence between 30 and 50 mas are expected, which can be probed by the interferometric diffraction limit of the LBT.

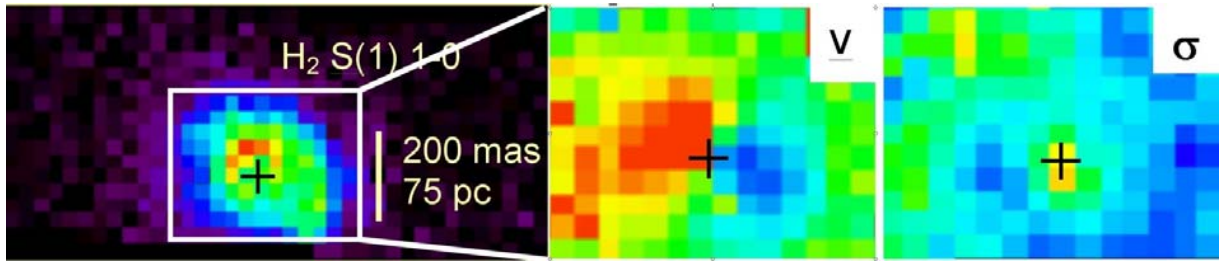

Fig. 2 Molecular hydrogen (right) around the center of NGC1275 observed with the imaging NIR spectrograph OSIRIS at Keck. The middle and right panel show the zoomed rotation and dispersion of that emission line from which the mass of the supermassive black hole can be inferred [7].

In addition, novel observation programs will become possible on M31 using LIINUS/SERPIL. The proper motions of the stars can be followed over several years, allowing for a direct measurement of the black hole mass, quite in the same way as has been done for the center of our Galaxy. The far greater distance to M31 is compensated for by the much higher velocity dispersion in the disk around its black hole.

## 2.3 Galactic Center

The black hole in the center of the Milky Way is the ideal laboratory for studying super-massive black hole. Recent observations using the SINFONI integral field spectrograph allowed to measure the radial velocities of stars near to SgrA* and to determine their 3-dimensional orbits [8]. This provides the most accuarate distance and mass estimate for the Galactic Center black hole. Fig. 3 shows the first adaptive optics integral field observations of the Galactic Center, [9] performed with SINFONI.

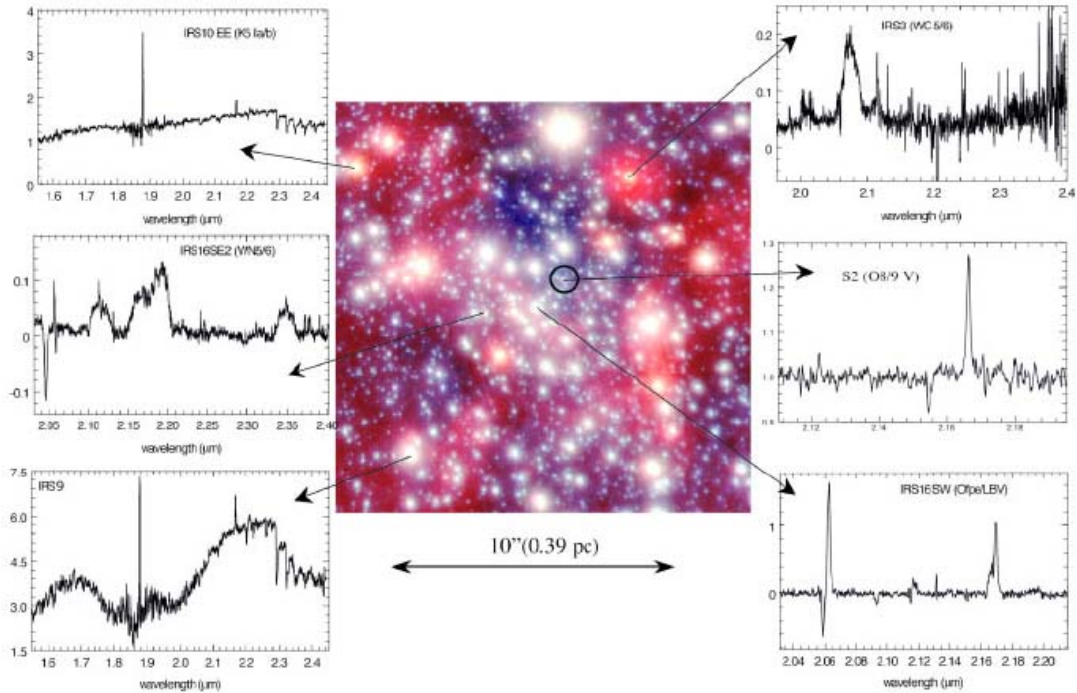

Fig. 3. Center of the Milky Way observed with the imaging spectrograph SINFONI. The improvement of the spatial resolution is required to be able to resolve the very dens regions of the galactic center.

Another hot topic is the flare activity of SgrA*. Since these flares are weak and strongly blended with light from neighboring stars (see e.g. Fig. 4) the flare's physics can only be disentangled using instruments offering spatial resolutions better than a few tens of milliarcseconds.

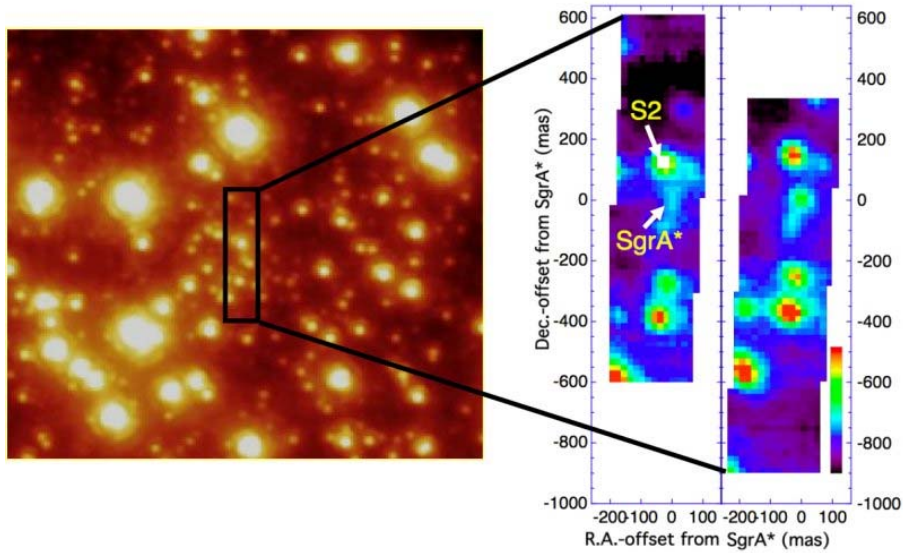

Fig. 4. NIR image of the Galactic Center (left) taken by [10] and collapsed OSIRIS K-band datacubes (right) from [11]. The two data cubes have been obtained consecutively during a period of activity of SgrA* with the new imaging near-infrared spectrograph OSIRIS at Keck observatory. The spatial resolution achieved during the laser assisted observations of SgrA* is about 50 mas, corresponding to 400 AU.

Imaging spectroscopy with LIINUS/SERPIL will allow to resolve the stellar content at factor 2-3 higher angular resolution, and probe the immediate vicinity of the central black hole at a distance of approximately 100 AU.

**2.4 Extrasolar Planets**

About 170 extrasolar planets have been detected so far, most of them indirectly, i.e. through periodic Doppler-shift of stellar absorption lines. However, these existing observations do hardly disclose any detailed information about their atmosphere and their formation. The increasing size and angular resolution of present telescopes offers the opportunity to conduct observations of extrasolar planets. The first direct detection of exoplanets was recently published [12, 13]. The observing strategy is depicted in Fig. 5. Although photometrical measurements with high angular resolution are currently being done by masking out the star with a cone, the spectral information obtained with an IFS is essential. LIINUS/SERPIL will be the proper instrument for such observations, since it provides the necessary spatial resolution. Although a more exhaustive exploration of exoplanets will be the task of the next generation 30-100m class telescopes, the LBT with its interferometric capability and an imaging spectrograph remains the only telescope for such observations in the near future.

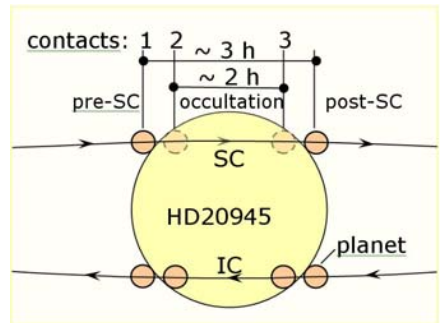

Fig. 5 The exoplanet moves across the stellar disk during inferior conjunction and disappear during superior conjunction. The difference between the spectrum of the star and the spectrum of the star after recurrence of the planet reveals the planetary spectrum [14].

## 3 INSTRUMENT CONCEPT OF LIINUS/SERPIL

This section describes the overall instrument concept including trade-off and system analysis, as well as the derived instrumental requirements.

**3.1 Instrument and technical requirements**

The specifications and design of LIINUS/SERPIL are based on detailed trade-offs between the various technical options of integral field spectrometers and are driven by the science requirements presented above. We have extracted a set of preliminary instrument specifications for the near-infrared interferometric integral field spectrograph in Table 1.

In addition, the complete spectrograph will be cooled to cryogenic temperatures. The detectors will be operated at about 77 K, but, depending on the instrument configuration, part of the instrument might work at ambient temperature.

Due to LBTs alt-azimuth mount, the sky rotates during observations with respect to the entrance pupil of the telescope. Most instruments, like LBC, LUCIFER, MODS, etc. are equipped with de-rotators, which permit long exposures without trailing. In case of LBT, the LINC-NIRVANA instrument must remain fixed with regard to the telescope pupil for proper interferometric operation. Actually, there are two options to deal with this difficulty: (1) allowing the sky to rotate will produce blurring for any objects that are not in the center of the field, and (2) rotating the detector will prevent trailing, but causes PSF rotation in place. The rate of field rotation depends on the declination and hour angle of the astronomical target, but it can result in image motion of many pixels in the corners of the field for modest exposure times. This means that the second option is far superior for any realistic scientific program, where long exposure times are indispensable. Therefore there was a detector de-rotation device incorporated into the cryostat.

Table 1. Preliminary high level science requirements for LIINUS/SERPIL

| Parameter | Value | Comments |
|---|---|---|
| Wavelength range | 1.1 – 2.4 micron | J, H, K infrared bands |
| FoV | 0.5 - 1 arcsec | Diameter in both directions |
| IFU spatial sampling | ≤30 x ≤10 mas | At the low and high resolution direction of LBT |
| Spectral sampling | 2 detector pixels | Nyquist sampling |
| Spectral Resolution | ≥ 3000 | OH line correction |
| Pixel size | 15 – 18 micron | |
| Pixel number | 2kx2k – 4kx4k | |
| Total system throughput | > 25 % | |

Tracking the sky automatically means that the entrance pupil rotates from the point of view of the detector. Thus, the interferometric PSF rotates on the detector around its center, which causes blurring of the image, as is shown in Fig. 6. The situation is mitigated somewhat by the small separation of the LBT mirrors compared to their diameter, which translates to a small number of fringes (3) across the core of the 8.4 m Airy disk – a sparse array would produce more such fringes and greater sensitivity to rotational blurring.

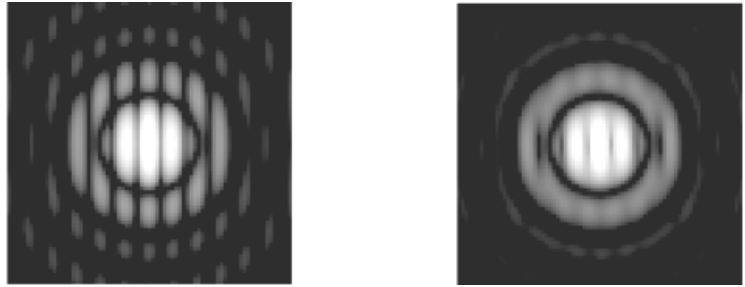

Fig. 6. Pupil signature of the PSF in case of a single exposure (left). For longer integration times the image of the PSF blurs due to the pupil rotation. The image on the right corresponds to about 30 degree of pupil rotation and illustrates the degradation of the angular resolution of about 10% along the horizontal center line.

Clearly, very long integrations or observations near the zenith, where the parallactic angle changes rapidly, can have a deleterious effect on the high spatial frequencies delivered by the instrument. For total integration times of the order of one hour the image rotation will not be an issue if the zenith region is avoided. The quantitative investigation of this problem is currently being completed.

**3.2 System architecture**

The principle of an Integral Field Spectrograph (IFS) is to realize a sampling of the field and to generate a spectrum of each field sample. An IFS consists of the pre-optics, which re-images the FoV focused by the telescope onto an image slicing device that is fed into a conventional spectrograph. Existing integral field spectrographs use a variety of techniques to deliver a spatially-resolved spectroscopic capability. For LIINUS/SERPIL, the following three techniques of Integral Field Units (IFU) are explored: microlenses, fiber bundles, and image slicers. The choice of the IFU for LIINUS/SERPIL depends on the space available in the LINC-NIRVANA cryostat. This space is about half of the volume between two coaxially placed cylinders. The radius of the smaller cylinder is 8 cm, the radius of the bigger one is 35 cm, and their height is 70 cm. There also is an option to fiber-fed an external cryostat, or to build a completely new cryostat with accordingly higher cost and effort. Placing the instrument inside LINC would be most desirable, if technically feasible, and satisfying the top-level requirements. The following solutions have been identified for LIINUS/SERPIL, displayed in Fig. 7.

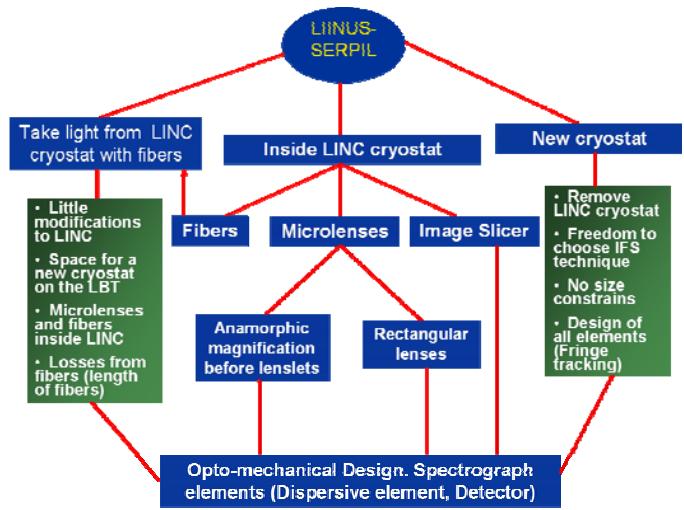

Fig. 7 Technical solutions for LIINUS/SERPIL

### 3.3 Image slicing devices

The design of the image slicing device is of fundamental importance concerning the space requirement we must meet. In the following, three types of image slicers are considered for the instrument: image slicers, microlens arrays, and fiber bundles.

### 3.3.1 Advanced image slicer

The slicer IFU solution uses mirrors in the focal plane to cut the field of view into a number of strips, which are then aligned one-by-one to form the pseudo-longslit that is fed into the spectrograph. This solution is more efficient in terms of spectral coverage and packing efficiency. An example of a successful working slicer IFU in operation today is SINFONI built by MPE and ESO [15, 16]. Due to the length of the pseudo-slit, these IFUs normally occupy a lot of space in the cryostat. This solution for LIINUS/SERPIL would mean that the instrument could probably not be built inside the LINC cryostat, implying the construction of a separate cryostat for the instrument, and the design of other subsystems such as the fringe and flexure tracking. A variation of this principle which minimizes the size of the pseudo-slit has been proposed by [17]; the mirrors have curved surfaces to allow the pupil-relay mirrors to be reduced in size. These types of IFUs are known as advanced image slicers. A reduction in size of the long slit by a factor of > 3 may be sufficient to implement LIINUS/SERPIL inside LINC. Another possibility for an advanced image slicer would be to place lenses in front of the second mirror which perform the reduction of the individual strips [18].

### 3.3.2 Microlenses

The most compact way of image slicing is employing microlens arrays. Such lenslet arrays provide the required high spatial sampling of the image and the overall size of the optical system remains relatively compact. Present technologies offer rectangular- or hexagonal-shaped lens arrays with a large variety of sizes and focal lengths, well matched to the need for LIINUS/SERPIL. Examples of existing instruments accommodating microlens arrays are OSIRIS [19] and TIGER [20] shown in Fig. 8. Designing rectangular-shaped microlenses as non-spherical lens arrays also could directly solve the imaging problem of anamorphic magnification (see below) needed for LIINUS/SERPIL, and their rectangular pupil would fit well the 15x15micron detector pixels.

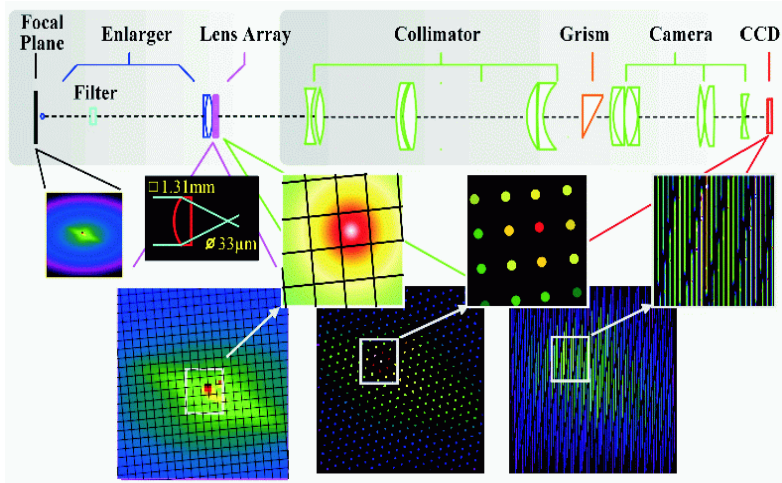

Fig. 8. Example of an integral field imaging spectrograph [20]. The IFS includes three sections: the telescope field is first enlarged in order to match the microlens array dimensions. The microlenses array performs the field sampling and focuses the incoming light on micropupils, which imaged and dispersed through a classical spectrograph with collimator, dispersing element and focusing optics onto a CCD camera.

A disadvantage of applying a microlens array is the problem of accommodation of the spectra on the detector array. The image of the micro pupils are placed rather densely on the detector (1-2 pixel between spectra), therefore the single spectra of a spatial elements may require more sophisticated software for proper extraction.

### 3.3.3 Microlenses applied with fibers

Another solution of image slicing is the combination of microlens arrays and optical fiber bundles. This approach would be the most flexible. It allows to feed the light to an external spectrograph, which would overcome the space-limitations in the LINC-NIRVANA cryostat, and would allow an optimum spectrograph design. With optical fibers the image slicing can be done in a similar fashion as with microlenses, but the arrangement of the spectral lines can be selected such as to fit the dimension of the detector best without making compromises on the spectral resolution. The major drawbacks of the fiber bundles are the coupling losses and the comparably high costs.

## 4 PRELIMINARY OPTICAL DESIGN USING MICROLENSES

The image slicing in this design is performed by a micro-lens array having 0.6 mm separation between the adjacent microlenses. This solution is investigated first, because the integral field spectrometer could completely fit inside the LINC cryostat. The specifications for the spectrograph in this design are: a field of 1 x 1 arcsec$^2$, a sampling of about 30x10 mas and a spectral resolution R= $\lambda/\Delta\lambda$ of $\geq 3000$.

### 4.1 Re-imaging optics

The field magnifier optics adapts the focal length of the telescope such that the 0.6 mm microlens pitch matches the field sampling at 0.03x0.01 arcsec. This requires an effective focal length of 12375 mm, corresponding to a magnification of 16.1 over the LINC focal plane. The beam speed is f/515 at the microlens array entrance. In addition, the optical beam needs to be telecentric in order to keep the mechanical dimension of the optics acceptable.

### 4.2 Anamorphic magnification

The light collected with LBT is coherently combined in the Fizeau interferometric mode, thus the resolution along the interferometric axis is about 3.3 times the resolution perpendicular to it. Since we are generally dealing with quadratic pixels, the pixels in the low resolution direction will be wasted. An effective use of the capabilities of the IFU should include anamorphic imaging. Consequently, increasing about threefold the field of view in the low resolution direction results in the same number of spatial pixel than in the high resolution direction, without making any compromises in the spatial resolution.

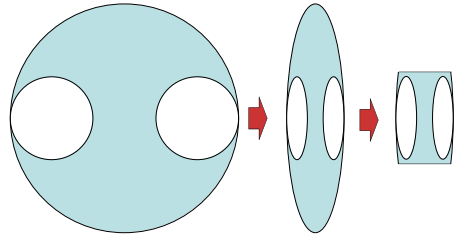

Fig. 9. Impact of the anamorphic magnification on the pupil geometry.

A differential rescaling by a factor of about 3 results in an elliptical pupil with the same axis ratio. The new geometry of the pupil is illustrated in Fig. 9. The figure also shows the geometry of the effective pupil, where the non-active areas have been blocked out. The effective size is almost a square, which helps the optics downstream. An illustrative representation of the anamorphic magnification is shown in Fig. 10, where an image of the sky is magnified by a factor of 3 then reconstructed according to the quadratic-shaped detector.

The anamorphic magnification in the optical chain can be realized in two ways. One option is that the microlens array is designed to have different focusing in the "vertical" and "horizontal" directions and the rest of the magnification needed is done by the pre-optics. The other possible solution is to place an additional imaging optics in front of the commercially used microlens array. The paraxial computation of a three lens anamorphic magnifier with non-spherical lenses, looks feasible. Ideally, the anamorphic magnifying optics will be integrated into the general magnification of the image scale (pre-optics)

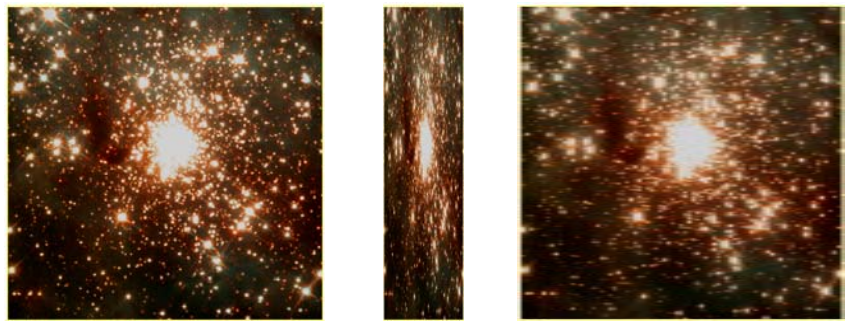

Fig. 10. Image samples through the optical system applying anamorphic magnification. The target object (left) is magnified in the vertical direction (middle) then reconstructed again (right). The loss of resolution in the horizontal direction is obvious.

As an alternative, an anamorphic prism pair is considered as well. However, such a set-up only allows for the differential magnification of about 3 and can not serve as enlarger optics. Hence, an additional overall magnification of a factor of 2.5 will be still necessary in this case to match the image scale.

### 4.3   Microlens array

The sampling of the field is performed by a 20 mm x 60 mm array of microlenses with a separation between centers of about 0.6 mm x 0.6 mm. With a hexagonal microlens array, the horizontal sampling with 100 hexagons of 0.6 mm pitch leads to a hexagon side of 346 micron. The interline distance or vertical sampling pitch is 520 micron leading to 38 lines. So the 1 arcsec square field of view is sampled with a 0.01 arcsec pitch horizontally and 0.03 arcsec vertically.

The output face of the microlens array is a pupil plane. An f-number of f/13 has been selected as reasonable compromise and the microlens focal length is thus 7.6 mm. The microlens pupil diameter results from the microlens focal length and beam f-number at microlens entrance and is equal to 15 micron. Considering diffraction effects this diameter increases to 30 micron.

### 4.4 Spectrograph section

The spectrograph section consists of a collimator, the dispersive element and a camera. These elements reimage and disperse the light from the micro-pupils on the detector. The pupil of the spectrograph section is determined by the shape of the microlenses. The spectrograph section has a magnification of 1 x 2 which results in a pupil image diameter of 30 micron i.e. about 2 pixels. The spectrograph design is adjusted to get a disperser size of 10 mm. The microlens system provides telecentric beams.

#### 4.4.1 Collimator

The collimator has a focal length of 127 mm. It was designed starting from the grating to the microlens array in order to obtain an image and evaluate the design by measuring the aberrations. An optimized dioptric collimator has been developed with telecentric beams at the exit of the microlens array. This collimator delivers diffraction-limited imagery over the 1.1-2.45 micron range and at an image scale of 10 mas/pixel. The field of view is 36º. The entrance pupil was chosen to be 20 mm in front of the first lens surface. The total length of the collimator is 250 mm.

#### 4.4.2 Dispersive element

The spectral resolution required for the spectrograph can be achieved using several well established techniques including transmission or reflecting diffraction gratings, grisms - a prism with a grating on one of its surface - and volume phased holographic (VPH) grating. We presently study a grism (depicted in Fig. 11) or VPH grating solution for LIINUS/SERPIL. Grisms have the advantage that the alignment is simple, significantly reduces imaging aberration of coma across the spectral range and defocus at the range edges. Furthermore, the prism itself contributes to the overall dispersion, which results in a smaller groove density then that of a simple grating. This is very important to avoid polarization effects caused by the high groove density and to improve the grating efficiency. Good results were achieved with the grism ruling if the grating constant is at least five times the wavelength [21]. The avoidance of polarization effects is very important in order to guarantee the good diffraction efficiency of the grating over the entire spectral range of 1.95-2.45 micron. Example of a spectrographs using a grism is 3D [22]. In this instrument, the grism is made of KRS-5, which has a high refractive index of about 2.4, good optical quality at the near-infrared wavelengths, and good mechanical properties, which is crucial for the ruling procedure. Recent developments of grisms employ photolithographic technology and surface etching of Silica, which also seems to provide the high accuracy of the grating profile In addition, Silica has an even larger refractive index than KRS-5, about 3.4 at 2.7 micron. Lab experiments are planned in order to optimize the grating performance and the technological processes.

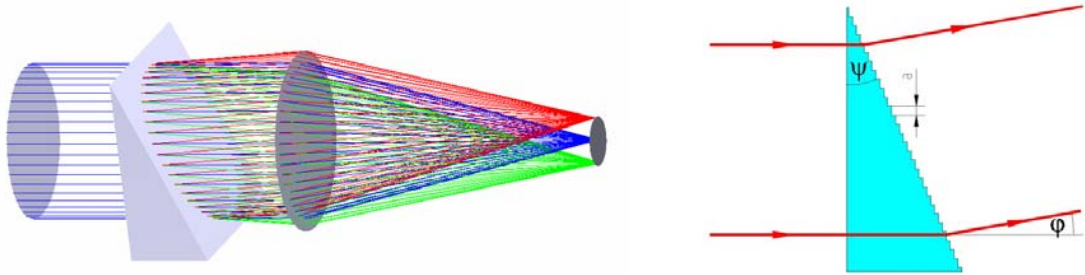

Fig. 11. Design of the LIINUS/SERPIL grism (left). By using higher diffraction orders (m=3, 4, 5) the bandwidth of the instrument will cover the entire J, H, and K bands. In this configuration, one specific wavelength in each band propagatesparallel to the optical axis. The profile of the grooves is sawtooth (right) to take advantage of the maximum grating efficiency.

Besides conventional gratings, VPH gratings are also considered in spectrographs used for astronomical applications [23]. The diffraction process in VPH gratings originates from modulations in the refractive index in the form of fringe planes running parallel to each other through the grating material and oriented such that the fringes terminate at the surfaces of the volume. The frequency of the fringes determines the grating dispersion and the diffraction geometry must fulfil the classical Bragg condition. The technology currently available provides high efficiency VPH gratings already at near-infrared wavelengths with standard line densities of 300 or 600 lines/mm, fitting well into the design of LIINUS/SERPIL.

### 4.4.3 Camera optics

The camera optics is considered for the instrument concepts using a 2kx2k or a 4kx4k detector. In the optimistic case, the 4kx4k detector will be available in the near future, thus investigation were carried out for the camera optics using this detector. The nominal focal length of the camera is 260 mm. The field of view is 19.2º and the entrance pupil of the camera is chosen to be 100 mm in front of the first lens surface (for the VPH grating solution). The camera can be designed to be completely diffraction limited at all wavelengths (1.1 – 2.45 micron) when operated with the interferometric image scale of 10 mas/pixel. The camera was optimized for the smallest spot size. The total length of the camera is 400 mm. The severe constrains on the image quality and the size of the instrument can better be encountered by using lenses made from $BaF_2$ and the special infrared glass IRG2 from Schott. This combination is known as one of the best combinations for lenses in the 1 -2.45 micron range [24]. These glasses were used for the design of the collimator and the camera.

### 4.4.4 Detector

LIINUS/SERPIL is designed for a 4k x 4k pixel array. This could be a mosaic of four 2k x 2k detectors (e.g., Rockwell HAWAII 2, or Ratheon Infrared Center of Excellence) with 18 micron pixels. For the design study, however, we assume that in the near future 4k x 4k arrays will be available, with a likely pixel size of 15 micron, yielding a total array physical size of 62 mm. Quantum efficiencies are expected to be greater than 80% for the near-infrared with a read noise of approximately 2 electrons. The distance in the focal plane between the pupil images is 1.22 mm or 80 pixels. The spectrum width is about 2 pixels. For optimum packing of the spectra, the dispersive element is slightly rotated with respect to the detector. Using a small angle, long spectra fit well, but the distance between spectra is reduced. A reasonable compromise is needed with a distance between 2 and 4 pixels between spectra.

## 5    SUMMARY AND CONCLUSIONS

We have introduced the concept LIINUS/SERPIL, a near-infrared integral field spectrograph for interferometric diffraction-limited observations at the LBT. We also explained part of the science that makes such an instrument very attractive. The state-of-the-art design as well as its optical and spectroscopic capabilities pushes LIINUS/SERPIL well beyond the limits of the current instrumentation at large telescopes.